\documentclass[runningheads]{llncs}
\usepackage{graphicx}
\usepackage[left=1in, right=1in, top=1in, bottom=1in]{geometry}
\usepackage{amsmath}
\usepackage{amsfonts}
\usepackage{amssymb}
\usepackage{hyperref}
\usepackage{booktabs}
\usepackage{multirow}
\usepackage[table,xcdraw]{xcolor}
\usepackage[numbers]{natbib}
\begin{document}
\title{Generalized Dice Focal Loss trained 3D Residual UNet for Automated Lesion Segmentation in Whole-Body FDG PET/CT Images}
%
%
\author{Shadab Ahamed\inst{1, 2}\orcidID{0000-0002-2051-6085} \and
Arman Rahmim\inst{1,2}\orcidID{0000-0002-9980-2403}}
\institute{University of British Columbia, Vancouver, BC, Canada \and
BC Cancer Research Institute, Vancouver, BC, Canada}
\maketitle              
\begin{abstract}
Automated segmentation of cancerous lesions in PET/CT images is a vital initial task for quantitative analysis. However, it is often challenging to train deep learning-based segmentation methods to high degree of accuracy due to the diversity of lesions in terms of their shapes, sizes, and radiotracer uptake levels. These lesions can be found in various parts of the body, often close to healthy organs that also show significant uptake. Consequently, developing a comprehensive PET/CT lesion segmentation model is a demanding endeavor for routine quantitative image analysis. In this work, we train a 3D Residual UNet using Generalized Dice Focal Loss function on the AutoPET challenge 2023 training dataset. We develop our models in a 5-fold cross-validation setting and ensemble the five models via average and weighted-average ensembling. On the preliminary test phase, the average ensemble achieved a Dice similarity coefficient (DSC), false-positive volume (FPV) and false negative volume (FNV) of 0.5417, 0.8261 ml, and 0.2538 ml, respectively, while the weighted-average ensemble achieved 0.5417, 0.8186 ml, and 0.2538 ml, respectively. Our algorithm can be accessed via this link: 
https://github.com/ahxmeds/autosegnet.

\end{abstract}

\section{Introduction}
\label{sec:introduction}
Fluorodeoxyglucose ($^{18}$F-FDG) PET/CT imaging is the gold standard in cancer patient care, offering precise diagnoses, robust staging, and valuable therapy response assessment \cite{pet_importance}. However, in the conventional approach, PET/CT images receive qualitative evaluations from radiologists or nuclear medicine physicians, which can introduce errors stemming from the inherent subjectivity among different expert readers. Incorporating quantitative assessments of PET images holds the promise of enhancing clinical decision-making precision, ultimately yielding improved prognostic, diagnostic, and staging outcomes for patients undergoing various therapeutic interventions \cite{tmtv, suvmean_tmtv_tlg}. \\

Quantitative evaluation often involves manual lesion segmentation from PET/CT images by experts, which is a time-consuming task and is also prone to intra- and inter-observer variabilities \cite{FOSTER201476}. Automating this task is thus necessary for routine clinical implementation of quantitative PET image analysis. Traditional thresholding-based automated techniques usually miss low-uptake disease and produce false positives in regions of physiological high uptake of radiotracers (such as the brain, bladder, etc.). To combat these limitations, deep learning offers promise for automating lesion segmentation, reducing variability, increasing patient throughput, and potentially aiding in the detection of challenging lesions, providing valuable support for healthcare professionals \cite{Wang2021, Tang2019, Hollon2020, Mei2020}.\\

While there have been significant strides in the field of deep learning-based segmentation, tackling the segmentation of lesions from PET/CT images remains a challenging task. This difficulty primarily stems from the scarcity of large, meticulously annotated publicly-accessible datasets. In the majority of research endeavors, deep learning models are trained on relatively modest, privately owned datasets. This limitation in data availability not only hampers models' generalizability but also hinders their widespread adoption for routine applications. Organizing data challenges like the AutoPET segmentation challenge \cite{Gatidis2022, autopet2022_paper}, which provides access to extensive, publicly available PET/CT datasets, marks a significant milestone in advancing the development of highly accurate models capable of meeting the stringent criteria necessary for clinical implementation. The AutoPET challenge dataset stands out for its remarkable diversity, encompassing patients with various cancer types, including lymphoma, lung cancer, and melanoma alongside a group of negative control patients. This diverse composition enhances the dataset's representativeness and broadens its potential applications in the medical field. \\

In this work, we trained a 3D Residual UNet using the training dataset (1014 PET/CT pairs and ground truth segmentation masks from 900 patients) provided in the challenge. Testing was first performed by submitting the trained algorithm to the challenge preliminary test phase which consisted of 5 cases. The algorithm was then submitted to the final test phase which consisted of 200 cases for final evaluation.

\section{Materials and Methods}
\label{sec:materials_and_methods}
\subsection{Data and data split}
\label{subsec:data_and_datasplit}
The training data consisted of four cohorts, namely scans presenting lymphoma (145 cases), lung cancer (168 cases), melanoma (188 cases), and negative control patients (513 cases). Each of these 4 cohorts was individually split into 5 folds. The final 5 training folds were created by segregating together the images belonging to fold $f$ across the four cohorts, where $f = \{0, 1, 2, 3, 4\}$. No other dataset (public or private) was used in this work.   

\subsection{Preprocessing and data augmentation}
\label{subsec:preprocessing_and_data_augmentation}
The CT images were first downsampled to match the coordinates of their corresponding PET images. The PET intensity values in units of Bq/ml were decay-corrected and converted to SUV. During training, we employed a series of non-randomized and randomized transforms to augment the input to the network. The non-
randomized transforms included (i) clipping CT intensities in
the range of [-1024, 1024] HU (ii) min-max normalization of clipped CT intensity to
span the interval [0, 1], (iii) cropping the region outside the body in PET, CT, and mask images using a 3D bounding box, and (iv) resampling the PET, CT, and mask images to an isotropic voxel spacing of (2.0 mm, 2.0 mm, 2.0 mm) via
bilinear interpolation for PET and CT images and nearest-neighbor interpolation for mask images. \\

On the other hand, the randomized transforms were called at the start of every epoch. These included (i) random spatial cropping of cubic patches of dimensions (192, 192, 192) from the images, (ii) 3D translations in the range (0, 10) voxels along all three directions, (iii) axial rotations by angle $\theta \in (-\pi/12, \pi/12)$, (iv) random scaling by a factor of 1.1 in all three directions, (v) 3D elastic deformations using a Gaussian kernel with standard deviation and offsets on the grid uniformly sampled from (0, 1), (vi) Gamma correction with $\gamma \in (0.7, 1.5)$, and (vii) addition of random Gaussian noise with $\mu = 0$ and $\sigma=1$. Finally, the augmented PET and CT patches were concatenated along the channel dimension to construct the final input for the network.

\subsection{Network}
\label{subsec:network}
We used a 3D Residual UNet \cite{resunet}, adapted from the MONAI library \cite{monai_paper}. The network consisted of 2 input channels, 2 output channels, and 5 layers of encoder and decoder (with 2 residual units per block) paths with skip-connections. The data in the encoder
was downsampled using strided convolutions, while the decoder unsampled using transpose strided convolutions. The number of channels in the encoder part from the top-most layer to the bottleneck were 32, 64, 128, 256, and 512. PReLU was used as the activation function within the network. The network consisted of 19,223,525 trainable parameters.

\subsection{Loss function, optimizer, and scheduler}
\label{subsec:loss_function_optimizer_and_scheduler}
We employed the binary Generalized Dice Focal Loss $\mathcal{L}_{\text{GDFL}} = \mathcal{L}_\text{GDL} + \mathcal{L}_\text{FL}$, where $\mathcal{L}_\text{GDL}$ is the  Generalized Dice Loss \cite{gendiceloss} and $\mathcal{L}_\text{FL}$ is the Focal Loss \cite{focalloss}. The Generalized Dice Loss $\mathcal{L}_\text{GDL}$ is given by,
\begin{equation}
 \mathcal{L}_\text{GDL} = 1 - \frac{1}{n_b} \sum_{i=1}^{n_b} \frac{\sum_{l=0}^{1} w_l \sum_{j=1}^{N^3} p_{ilj} g_{ilj}  + \epsilon}{\sum_{l=0}^{1} w_l \sum_{j=1}^{N^3}(p_{ilj} + g_{ilj})  + \eta}
\end{equation}
where $p_{ilj}$ and $g_{ilj}$ are values of the $j^{th}$ voxel of the $i^{th}$ cropped patch of the predicted and ground truth segmentation masks with class $l \in \{0, 1\}$ respectively in a mini-batch size $n_b$ of the cropped patches and $N^3$ represents the total number of voxels in the cropped cubic patch of size $(N, N, N)$, where $N=192$. Here, $w_l = 1/(\sum_{j=1}^{N^3}g_{ilj})^2$ represents the weight given to class $l$. The mini-batch size was set to $n_b=4$. Small constants $\epsilon = \eta = 10^{-5}$ were added to the numerator and denominator, respectively to ensure numerical stability during training.  The Focal Loss $\mathcal{L}_\text{FL}$ is given by,
\begin{equation}
    \mathcal{L}_\text{FL} = -\frac{1}{n_b}\sum_{i=1}^{n_b} \sum_{l=0}^{1} \sum_{j=1}^{N^3} v_l (1 - \sigma(p_{ilj}))^\gamma g_{ilj} \text{log}(\sigma(p_{ilj}))
\end{equation}
where, $v_0 = 1$ and $v_1 = 100$ are the focal weights of the two classes, $\sigma(x) = 1/(1 + \text{exp}(-x))$ is the sigmoid function, and $\gamma = 2$ is the focal loss parameter that suppresses the loss for the class that is easy to classify.\\

$\mathcal{L}_{\text{GDFL}}$ was optimized using the Adam optimizer. Cosine annealing scheduler was used to decrease the learning rate from $1 \times 10^{-3}$ to $0$ in 300 epochs. The loss for an epoch was computed by averaging the $\mathcal{L}_{\text{GDFL}}$ over all batches. The model with the highest mean Dice similarity coefficient (DSC) on the validation fold $f$ was chosen for further evaluation, for all $f \in \{0, 1, 2, 3, 4\}$.
\subsection{Inference and postprocessing}
\label{subsec:inference_and_postprocessing}
For the images in the validation set, we employed only the non-randomized transforms. The prediction was made directly on the 2-channel (PET and CT) whole-body images using a sliding-window technique with a window of dimensions $(192, 192, 192)$ and overlap=0.5. For final testing, the outputs of the 5 best models (obtained from 5-folds training) were ensembled via average and weighted average ensembling to generate the output mask. For the weighted average ensembling, the weights were chosen as the value of mean DSC of the respective validation fold. The final output masks were resampled to the coordinates of the original ground truth masks for computing the evaluation metrics.

\subsection{Evaluation metrics}
The challenge employed three evaluation metrics, namely the mean DSC, mean false positive volume (FPV) and mean false negative volume (FNV). For a foreground ground truth mask $G$ containing $L_g$ disconnected foreground segments (or lesions) $\{G_1, G_2, ..., G_{L_g}\}$ and the corresponding predicted foreground mask $P$ with $L_p$ disconnected foreground segments $\{P_1, P_2, ..., P_{L_p}\}$, these metrics are defined as,    
\begin{equation}
\label{eqn:dicescore}
    \text{DSC} = 2\frac{|G \cap P|}{|G| + |P|}
\end{equation}

\begin{equation}
\label{eqn:fpv}   
    \text{FPV} = v_p \sum_{l=1}^{L_p} |P_l| \delta(|P_l \cap G|) 
\end{equation}

\begin{equation}
\label{eqn:fnv}  
    \text{FNV} = v_g \sum_{l=1}^{L_g} |G_l| \delta(|G_l \cap P|) 
\end{equation}
where $\delta(x):= 1$ for $x=0$ and $\delta(x):= 0$ otherwise. $v_g$ and $v_p$  represent the voxel volumes (in ml) for ground truth and predicted mask, respectively (with $v_p = v_g$ since the predicted mask was resampled to the original ground truth coordinates). The submitted algorithms were ranked separately for each of the three metrics and the final ranking was determined based on the formula: $0.5 \times \text{rank}_\text{DSC} + 0.25  \times  \text{rank}_\text{FPV} + 0.25  \times \text{rank}_\text{FNV}$. The function definitions for these metrics were obtained from the challenge GitHub page and can be accessed via this \href{https://github.com/lab-midas/autoPET/blob/master/val_script.py}{link}.

\section{Results}
\label{sec:results}
We report the performance of our 5 models on the 5 validation folds in Table \ref{tab:validation_folds_results}. We have reported the mean and median values of the three metrics on the respective validation folds along with standard deviation and inter-quartile range respectively. On the preliminary test set, the average ensemble obtained DSC, FPV, and FNV of 0.5417, 0.8261 ml, and 0.2538 ml, respectively, while the weighted average ensemble obtained 0.5417, 0.8186 ml, and 0.2538 ml, respectively.

\begin{table}[h]
\caption{Network 5-fold cross-validation}
\centering
\resizebox{\textwidth}{!}{%
\begin{tabular}{|c|cccccc|}
\hline
 &
  \multicolumn{6}{c|}{\textbf{Metrics}} \\ \cline{2-7} 
 &
  \multicolumn{2}{c|}{\textbf{DSC}} &
  \multicolumn{2}{c|}{\textbf{FPV (ml)}} &
  \multicolumn{2}{c|}{\textbf{FNV (ml)}} \\ \cline{2-7} 
\multirow{-3}{*}{\textbf{\begin{tabular}[c]{@{}c@{}}Validation \\ Fold\end{tabular}}} &
  \multicolumn{1}{c|}{\textbf{Mean}} &
  \multicolumn{1}{c|}{\textbf{Median}} &
  \multicolumn{1}{c|}{\textbf{Mean}} &
  \multicolumn{1}{c|}{\textbf{Median}} &
  \multicolumn{1}{c|}{\textbf{Mean}} &
  \textbf{Median} \\ \hline
0 &
  \multicolumn{1}{c|}{0.61 ± 0.26} &
  \multicolumn{1}{c|}{0.71 {[}0.55, 0.82{]}} &
  \multicolumn{1}{c|}{4.16 ± 7.79} &
  \multicolumn{1}{c|}{1.41 {[}0.31, 3.95{]}} &
  \multicolumn{1}{c|}{10.92 ± 39.34} &
  0.31 {[}0.0, 5.41{]} \\ \hline
1 &
  \multicolumn{1}{c|}{\cellcolor[HTML]{FFFFFF}{\color[HTML]{212529} 0.62 ± 0.27}} &
  \multicolumn{1}{c|}{0.73 {[}0.49, 0.82{]}} &
  \multicolumn{1}{c|}{6.02 ± 14.13} &
  \multicolumn{1}{c|}{1.32 {[}0.24, 5.61{]}} &
  \multicolumn{1}{c|}{5.13 ± 11.53} &
  0.16 {[}0.0, 3.73{]} \\ \hline
2 &
  \multicolumn{1}{c|}{0.64 ± 0.25} &
  \multicolumn{1}{c|}{0.71 {[}0.52, 0.83{]}} &
  \multicolumn{1}{c|}{3.88 ± 7.94} &
  \multicolumn{1}{c|}{0.84 {[}0.19, 3.77{]}} &
  \multicolumn{1}{c|}{8.19 ± 20.93} &
  1.0 {[}0.0, 7.61{]} \\ \hline
3 &
  \multicolumn{1}{c|}{0.63 ± 0.27} &
  \multicolumn{1}{c|}{0.72 {[}0.53, 0.83{]}} &
  \multicolumn{1}{c|}{7.56 ± 13.79} &
  \multicolumn{1}{c|}{2.0 {[}0.52, 7.65{]}} &
  \multicolumn{1}{c|}{7.34 ± 21.23} &
  0.15 {[}0.0, 4.63{]} \\ \hline
4 &
  \multicolumn{1}{c|}{0.63 ± 0.26} &
  \multicolumn{1}{c|}{0.74 {[}0.48, 0.82{]}} &
  \multicolumn{1}{c|}{6.37 ± 12.37} &
  \multicolumn{1}{c|}{1.74 {[}0.41, 6.33{]}} &
  \multicolumn{1}{c|}{11.6 ± 43.44} &
  0.34 {[}0.0, 4.98{]} \\ \hline
\end{tabular}%
}
\label{tab:validation_folds_results}
\end{table}

\begin{table}[h]
\caption{Results of the preliminary test phase}
\centering
\resizebox{0.5\textwidth}{!}{%
\begin{tabular}{|c|c|c|c|}
\hline
\textbf{Ensemble} & \textbf{DSC} & \textbf{FPV (ml)}              & \textbf{FNV (ml)} \\ \hline
Average           & 0.5417       & 0.8261                         & 0.2538            \\ \hline
Weighted Average  & 0.5417       & \cellcolor[HTML]{FFFFFF}0.8186 & 0.2538            \\ \hline
\end{tabular}%
}
\label{tab:prelim_results}
\end{table}

\section{Conclusion and discussion}
\label{conclusion_and_discussion}
In this work, we trained a deep Residual UNet optimized via Generalized Dice Focal loss in a 5-fold cross validation setting to segment lesions from PET/CT images. Future work will include incorporating FPV and FNV terms into the loss function instead of just a Dice-based loss so that the predictions with high FPV or FNV can be particularly penalized.   

\section{Acknowledgment}
We appreciate the generous GPU support from Microsoft Azure, made available through the Microsoft AI for Good Lab in Redmond, WA, USA. We also acknowledge the valuable initial discussions with Sandeep Mishra.

\end{document}